%% file: ISMIR2022_template.tex
\documentclass{article}
\usepackage[T1]{fontenc} 
\usepackage[utf8]{inputenc} 
\usepackage{ismir,amsmath,cite,url}
\usepackage{graphicx}
\usepackage{color}
\usepackage{graphicx}
\usepackage{algorithm}
\usepackage{algpseudocode}
\usepackage{longtable}
\usepackage{rotating}
\usepackage{multirow}
\usepackage{booktabs}
\usepackage{makecell}
\usepackage{url}
\usepackage{enumitem}
\usepackage[toc,page]{appendix}

\title{Jointist: Joint Learning for Multi-instrument Transcription and Its Applications}

\multauthor
{Kin Wai Cheuk$^{*1,3}$ \thanks{$^*$The author conducted this work as an intern at ByteDance.} \hspace{1cm} Keunwoo Choi$^2$ \hspace{1cm} Qiuqiang Kong$^4$ \hspace{1cm} Bochen Li$^4$ \hspace{1cm}}{ 
\bfseries{ Minz Won$^4$ \hspace{1cm} Amy Hung$^4$ \hspace{1cm} Ju-Chiang Wang$^4$, \hspace{1cm} Dorien Herremans$^1$}\\
$^1$ Singapore University of Technology and Design \hspace{1cm}
$^2$ Gaudio Lab \\
$^3$ Agency for Science, Technology and Researc, Singapore \hspace{1cm}
$^4$ ByteDance\\
{\tt\small kinwai\_cheuk@mymail.sutd.edu.sg,  k@gaudiolab.com}
}

\begin{document}

\maketitle
\begin{abstract}
In this paper, we introduce Jointist, an instrument-aware multi-instrument framework that is capable of transcribing, recognizing, and separating multiple musical instruments from an audio clip.
Jointist consists of the instrument recognition module that conditions the other modules: the transcription module that outputs instrument-specific piano rolls, and the source separation module that utilizes instrument information and transcription results. 
%
The instrument conditioning is designed for an explicit multi-instrument functionality while the connection between the transcription and source separation modules is for better transcription performance. 

Our challenging problem formulation makes the model highly useful in the real world given that modern popular music typically consists of multiple instruments. However, its novelty necessitates a new perspective on how to evaluate such a model. During the experiment, we assess the model from various aspects, providing a new evaluation perspective for multi-instrument transcription. We also argue that transcription models can be utilized as a preprocessing module for other music analysis tasks. In the experiment on several downstream tasks, the symbolic representation provided by our transcription model turned out to be helpful to spectrograms in solving downbeat detection, chord recognition, and key estimation.

\end{abstract}
\section{Introduction}\label{sec:introduction}

Transcription, or automatic music transcription (AMT), is a music analysis task that aims to represent audio recordings as symbolic notations such as scores or MIDI (Musical Instrument Digital Interface) files \cite{benetos2013automatic, benetos2018automatic, piszczalski1977automatic}. 
AMT can play an important role in music information retrieval (MIR) systems since symbolic information -- e.g., pitch, duration, and velocity of notes -- determines a large part of our musical perception, distinguishing itself from other musical information such as timbre and lyrics.
A successful AMT can ease the difficulty of many MIR tasks by providing a denoised version of music in a musically-meaningful, symbolic format. There are examples that include melody extraction~\cite{ozcan2005melody}, chord recognition~\cite{wu2018automatic}, beat tracking~\cite{vogl2017drum}, composer classification \cite{kong2020large}, \cite{kim2020deep}, and emotion classification \cite{chou2021midibert}.
Finally, high-quality AMT systems can be used to build large-scale datasets as done in ~\cite{kong2020giantmidi}. This would, in turn, accelerate the development of neural network-based music composition systems as these are often trained using symbolic data and hence require large-scale datasets~\cite{brunner2018midi,wu2020transformer,hawthorne2018enabling}.
However, music transcription is a tedious task for human and a challenging task even for experienced musicians. As a result, large-scale symbolic datasets of pop music are scarce, impeding the development of MIR systems that are trained using symbolic music representations.

In early research on AMTs, the problem is defined narrowly: transcription is done for a single target instrument, which is usually piano \cite{klapuri1998automatic} or drums \cite{paulus2003model}, and whereby the input audio only includes that instrument. The limitation of this strong and then-unavoidable assumption is clear: the model would not work for modern pop music, which occupies a majority of the music that people listen to. In other words, to handle realistic use-cases of AMT, it is necessary to develop a multi-instrument transcription system. 
%
Recent examples are Omnizart \cite{wu2021omnizart, wu2020multi} and MT3~\cite{gardner2021mt3} which we will discuss in Section~\ref{subsec:miamt}.

\begin{figure*}[t]
  \centering
  \centerline{\includegraphics[width=14cm]{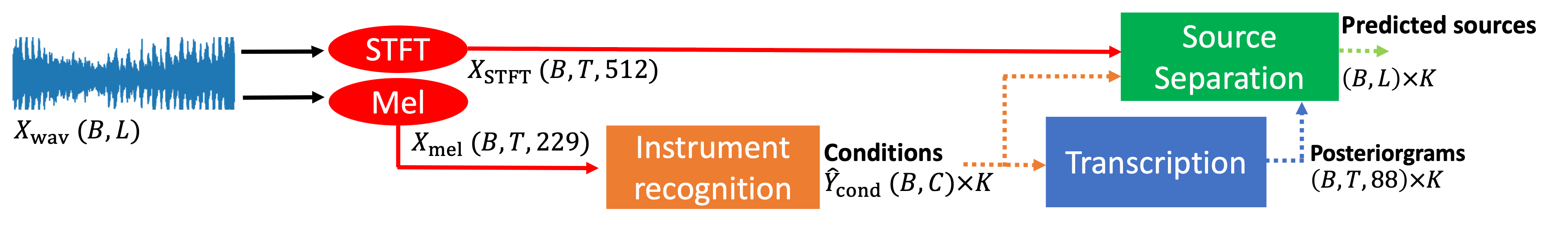}}
  \caption{The proposed Jointist framework. Our actual framework can transcribe/separate up to 39 different instruments as defined in Table~\ref{tab:MIDI_map} of Appendix. $B$: batch size, $L$: audio length, $C$: instrument classes, $T$: number of time steps, $K$: number of predicted instruments. Dotted lines represent iterative operations for $K$ times. Best viewed in color.}
  \label{fig:jointist}
\end{figure*}

In fact, the progress towards multi-instrument transcription has just begun, leaving several challenges related to the development and evaluation of such systems.
In particular,
the number of instruments in multiple-instrument audio recordings are not fixed. The number of instruments in a pop song may vary from a few to over ten. Therefore, it is limiting to have a model that transcribes a pre-defined fixed number of musical instruments in every music piece. Rather, a model that can adapt to varying number of target instrument(s) would be more robust and useful. This indicates that we may need to consider instrument recognition and instrument-specific behavior, during  development as well as evaluation. 


Motivated by the aforementioned recent trend and the existing issues,
we propose Jointist -- a framework that includes instrument recognition, source separation, and transcription. We adopt a joint training scheme to maximize the performance of transcription. 
First, 
we choose an ambitious scope of the task by including instrument recognition as a sub-task of the transcription and evaluating the performance of every existing instrument. This thorough evaluation would help researchers to deepen the understanding of the proposed framework -- its properties, merits, and limitations.
Second,
we adopt a fully supervised training scheme and rely on an existing dataset. This means that the target instrument set is determined by the dataset we use. This is a practical choice, although we hope that few-shot and zero-shot learning can alleviate this issue in the future.
Third, we provide experiment results that demonstrate the utility of transcription models as a pre-processing module of MIR systems. The result strengthens a perspective of transcription result as a symbolic representation, something distinguished from typical i) audio-based representation (spectrograms) or ii) high-level features \cite{choi2017transfer, castellon2021codified}. 


This paper is organized as follows. We first provide a brief overview of the background of automatic music transcription in Section~\ref{sec:background}. Then we introduce our framework, Jointist, in Section~\ref{sec:Jointist}. After that, we describe the experimental details in Section~\ref{section:experiments} and discuss the experimental results in Section~\ref{sec:discussion}. We also provide the experiment results and discussion on the application of Jointist in Section~\ref{sec:applications}. Finally, we conclude the paper in Section~\ref{sec:conclusion}.
\section{Background}
\label{sec:background}

\subsection{Multi-Instrument Automatic Music Transcription}
\label{subsec:miamt}
While automatic music transcription (AMT) models for piano music are well developed and are able to achieve a high accuracy~\cite{benetos2018automatic,Sigtia2015AnEN,kim2019adversarial,kelz2019deep,hawthorne2017onsets,hawthorne2018enabling,kong2021high}, multi-instrument automatic music transcription (MIAMT) is relatively unexplored. MusicNet~\cite{Thickstun2016LearningFO, Thickstun2017InvariancesAD} and ReconVAT~\cite{cheuk2021reconvat} are MIAMT systems that transcribe musical instruments other than piano, but their output is a \textit{flat} piano roll that includes notes from all the instruments in a single channel. In other words, they are not instrument-aware. Omnizart~\cite{8682605,wu2021omnizart} is instrument-aware, but it does not scale up well when the number of musical instruments increases as discussed in Section~\ref{sec:discussion_ss}. MT3~\cite{gardner2021mt3} is the current state-of-the-art MIAMT model. It formulates AMT as a sequence prediction task where the sequence consists of tokens of musical note representation. By adopting the structure of an NLP model called T5~\cite{raffel2020exploring}, MT3 shows that a transformer architecture can perform successful transcription by learning from multiple datasets for various instruments.    

\subsection{AMT and Joint Learning}
There have been attempts to jointly train a transcription model together with a source separation model. For example, \cite{jansson2019joint} uses the F0 estimation to guide the singing voice separation. However, they only demonstrated their method with a monophonic singing track. In this paper, the Jointist framework extends this idea into polyphonic music. While \cite{manilow2020simultaneous} extends the joint transcription and source separation training into polyphonic music, their model is limited to up to five sources (Piano, Guitar, Bass, Drums, and Strings) which do not cover the diversity of real-world popular music; and ignores the instrument-aware adaptation that the Jointist performs. 
While \cite{hung2021transcription} and \cite{chen2021zero} also use joint transcription and source separation training for a small number of instruments, they only use transcription as an auxiliary task, and there is no transcription during the inference phases.
On the contrary, \cite{tanaka2020multi} applies a joint spectrogram and pitchgram clustering method to improve the multi-instrument transcription accuracy, their model is only capable of doing transcription. A zero-shot transcription and separation model was proposed in \cite{lin2021unified} but was only trained and evaluated on 13 classical instruments.

\section{Jointist}\label{sec:Jointist}
We designed the proposed framework, Jointist, to jointly train music transcription and source separation modules so as to improve the performance for both tasks. By incorporating an instrument recognition module, our framework is instrument-aware and compatible with up to 39 different instruments.

As illustrated in Figure~\ref{fig:jointist}, Jointist consists of three sub-modules: the instrument recognition module $f_{\text{IR}}$, the transcription module $f_{\text{T}}$, and the music source separation module $f_{\text{MSS}}$. $f_{\text{IR}}$ and $f_{\text{T}}$ share the same mel spectrogram $x_{\text{mel}}$ as the input, while $f_\text{MSS}$ uses the STFT spectrogram $x_\text{STFT}$ as the input. 

Jointist is a flexible framework that can be trained end-to-end or individually. For ease of evaluation, we trained $f_{\text{IR}}$ individually, and we trained $f_{\text{T}}$ and $f_{\text{MSS}}$ together. In one of our ablation studies (Table~\ref{tab:ablation_source}), we also studied the model performance when training $f_{\text{T}}$ and $f_{\text{MSS}}$ separately.

\subsection{Instrument Recognition Module $f_\text{IR}$}
Transformer networks~\cite{vaswani2017attention} have been shown to work well for a wide range of MIR tasks~\cite{chen2021pix2seq, lu2021spectnt, huang2018music,park2019bi, won2019toward,guo2022musiac, won2021semi}. 
In this paper, we adopt the music tagging transformer proposed in~\cite{won2021semi} as our musical instrument recognition module,  $f_{\text{IR}}$.
Similar to~\cite{won2021semi}, our instrument recognition model consists of a convolutional neural network (CNN) front end and a transformer back end. The CNN front end has six convolutional blocks with output channels $[64, 128, 256, 512, 1024, 2048]$.
Each convolutional block has two convolutional layers with kernel size (3,3), stride (1,1), and padding (1,1), followed by average pooling (2,2).
To prevent overfitting, we perform dropout after each convolutional block with a rate of $0.2$~\cite{srivastava2014dropout}.
We conduct an ablation study as shown in Table~\ref{tab:inst} to find the optimal number of transformer layers, which turns out to be 4. 
The instrument recognition loss is defined as $L_\text{IR}=\text{BCE}(\hat{Y}_\text{cond},Y_\text{cond})$ where BCE is the binary cross-entropy, $\hat{Y}_\text{cond}$ is a tensor with the predicted instruments, and $Y_\text{cond}$ is a tensor with the ground truth labels.

\subsection{Transcription Module $f_\text{T}$}
Our transcription module $f_\text{T}$ is largely inspired by the model design of the onsets-and-frames model proposed in~\cite{hawthorne2017onsets}. We modify this module to be conditioned by instrument vectors using FiLM~\cite{perez2018film}, as proposed in~\cite{meseguer2019conditioned} for source separation. By performing this modification, we have a model that is capable of transcribing different musical instruments based on a given one-hot vector $I^i_\text{cond}$ of instrument $i$ defined in Table~\ref{tab:MIDI_map}. During inference, we combine the onset rolls and the frame rolls based on the idea proposed in~\cite{kong2021high} to produce the final transcription. In this procedure, the onset rolls are used to filter out noisy frame predictions, resulting in a cleaner transcription output.

$f_\text{T}$ consists of a batch normalization layer \cite{ioffe2015batch}, a frame model $f_\text{T}^\text{frame}$, and an onset model $f_\text{T}^\text{onset}$. The outputs from $f_\text{T}^\text{frame}$ and $f_\text{T}^\text{onset}$ are concatenated together and then passed to a biGRU layer followed by a fully connected layer (FC) to produce the final transcription $\hat{Y}_\text{T}$~\cite{chung2014empirical}. 
Teacher-forced training is used, i.e. the ground truth instrumental one-hot vectors $I^i_\text{cond}$ are used during training. During inference, $f_\text{IR}$ is used to generate the instrumental condition $\hat{Y}_\text{cond}$. The transcription loss is defined as $L_\text{T}=\sum_\text{j}L_j$, where $L_i = \text{BCE}(\hat{Y}_j, Y_j)$ is the binary cross entropy and $j\in\{\text{onset},\text{frame}\}$.


\subsection{Source Separation Module $f_\text{MSS}$}
We adopt a similar model architecture as in~\cite{jansson2017singing,meseguer2019conditioned} for our source separation module $f_\text{MSS}$. In addition to the the musical instrument condition $I^i_\text{cond}$ as the condition (either generated by $f_\text{IR}$ or provided by human users), our $f_\text{MSS}$ also uses the predicted piano rolls $\hat{Y}_T^i$ of the music as an extra condition. The source separation loss $L_\text{MSS}=\text{L2}(\hat{Y}_S^i,Y_S^i)$ is set as the L2 loss between the predicted source waveform $\hat{Y}_S^i$ and the ground truth source waveform $Y_S^i$. When combining the transcription output $\hat{Y}^i_\text{T}$ with $X_\text{STFT}$,
we explored two different modes: summation $g(\hat{Y}^i_\text{T})+X_\text{STFT}$ and concatenation $g(\hat{Y}^i_\text{T})\oplus X_\text{STFT}$. Where $g$ is a linear layer that maps the 88 midi pitches $\hat{Y}^i_\text{T}$ to the same dimensions as the $X_\text{STFT}$.


\section{Experiments}\label{section:experiments}
\subsection{Dataset}
Jointist is trained using the Slakh2100 dataset~\cite{manilow2019cutting}. This dataset is synthesized from part of the Lakh dataset~\cite{raffel2016learning} by rendering MIDI files using a high-quality sample-based synthesizer with a sampling rate of 44.1 kHz. The training, validation, and test splits contain 1500, 225, and 375 pieces respectively. The total duration of the Slakh2100 dataset is 145 hours. The number of tracks per piece in Slakh2100 ranges from 4 to 48, with a median number of 9, making it suitable for multiple-instrument AMT. In the dataset, 167 different plugins are used to render the audio recordings.

Each plugin also has a MIDI number assigned to it, hence we can use this information to map 167 different plugins onto 39 different instruments as defined in Table~\ref{tab:MIDI_map} in the Appendix.\footnote{\url{https://jointist.github.io/Demo/appendix.pdf}}
Our mapping is finer than the MIDI Instrument definition but slightly rougher than MIDI Programs.
For example, MIDI numbers 0-3 are different piano types. While MIDI instrument treats all of the them as Piano, the MIDI Program considers them as different instruments.
In this case, we believe that the MIDI Program is too fine for our task, hence we follow the MIDI Instrument mapping.
However, in another example, MIDI numbers 6 and 7 are harpsichord and clarinet, which sound totally different from the piano. Yet, they are both considered as piano according to the MIDI Instrument. In this case, we follow the MIDI Program and consider them as different instruments from the piano.
Finally, the original MIDI numbers only cover 0-127 channels. We add one extra channel (128) to represent drums so that our model can also transcribe this instrument.

\subsection{Training}
For all of the sub-modules of the Jointist framework,
the audio recordings are resampled to 16~kHz, which is high enough to capture the fundamental frequencies as well as some harmonics of the highest pitch $ C_{8} $ of the piano (4,186~Hz) \cite{hawthorne2017onsets, kong2021high}.
Following some conventions on input features~\cite{hawthorne2017onsets,kong2021high,cheuk2021reconvat}, for the instrument recognition and transcription s, we use log~mel~spectrograms -- with a window size of 2,048 samples, a hop size of 160 samples, and 229 mel filter banks. For the source separation, we use STFT with a window size of 1,028, and a hop size of 160. 
This configuration leads to spectrograms with 100 frames per second. Due to memory constraints, we randomly sample a clip of 10 seconds of audio from the full mix to train our models. 
The Adam optimizer \cite{kingma2014adam} with a learning rate of 0.001 is used to train both $f_\text{IR}$ and $f_\text{T}$. For $f_\text{MSS}$, the learning rate of $1\times10^{-4}$ is chosen after preliminary experiments.
All three sub-modules are trained on two GPUs with a batch size of six each. Pytorch and Torchaudio~\cite{paszke2019pytorch, yang2022torchaudio} are used to perform all the experiment and audio processing.

The overall training objective $L$ is a sum of the losses of the three modules, i.e., $L=L_\text{IR}+L_\text{T}+L_\text{MSS}$. 


\input{tables/instrument_recognition}

\input{tables/transcription}

\subsection{Evaluation}
We report the F1 scores for $f_\text{IR}$ with a classification threshold of 0.5 for all of the instruments. We chose this to ensure the simplicity of the experiment, even though the threshold can be tuned for each instrument to further optimize the metric \cite{musicclassification:book}. To understand the model performance under different threshold values, we also report mean average precision (mAP) as shown in Table~\ref{tab:inst}.

For $f_\text{T}$, we propose a new instrument-wise metric to better capture the model performance for multi-instrument transcription. Existing literature uses mostly flat metrics or piece-wise evaluation~\cite{hawthorne2017onsets,gardner2021mt3,kong2021high, cheuk2021reconvat}. Although this can provide a general idea of how good the transcription is, it does not show which musical instrument the model is particularly good or bad at. Since frame-wise metrics do not reflect the perceptual transcription accuracy~\cite{Cheuk_IJCNN2021}, we report only the note-wise (N.) and note-wise with offset (N\&O) metrics in Table~\ref{tab:transcription}.

For $f_\text{MSS}$, we also report the instrument-wise metrics to better understand the model performance (Table~\ref{tab:source_separation}).

\input{tables/source_separation}
\input{tables/ablation}

\section{Discussion}
\label{sec:discussion}
\subsection{Instrument Recognition}
Table~\ref{tab:inst} shows the mAP and F1 scores of the instrument recognition module $f_\text{IR}$ when different numbers of transformer encoder layers are used. Both the mAP and F1 scores improve as the number of layers increases. The best mAP and F1 scores are achieved when using four transformer encoder layers. Due to the instrument class imbalance in the Slakh2100 dataset, our $f_{IR}$ performance is relatively low for instrument classes with insufficient training samples such as clarinet, violin, or harp. Therefore, the weighted F1/mAP is higher than the macro F1/mAP.\footnote{The F1 score for each instrument is available in the supplementary material, \url{https://jointist.github.io/Demo/}}. Nevertheless, we believe that our $f_{IR}$ is good enough to generate reliable instrument conditions for popular instruments such as drums, bass, and piano.

\subsection{Transcription}
\label{sec:discussion_transcription}
When evaluating the transcription module $f_\text{T}$, we assume that we have a perfect $f_\text{IR}$ and use the ground truth instrument labels $Y_\text{cond}$ as the conditions for $f_\text{T}$. This is a choice to isolate our evaluation to the transcription module; otherwise, incorrect predictions of $f_\text{IR}$ will affect the evaluation of  $f_\text{T}$ part. 

Table~\ref{tab:transcription} shows the note-wise (N.) and note-wise with offset (N\&O) transcription F1 scores for different models.
In addition to training only the $f_\text{T}$ (Model T), we also explore the possibility of training both $f_\text{T}$ and $f_\text{MSS}$ jointly (Model T+S) and study its effect on the transcription accuracy.

When training $f_\text{T}$ standalone for 1,000 epochs, we can achieve an instrument-wise F1 score of 23.2. We presumed that joint training of $f_\text{T}$ and $f_\text{MSS}$ would result in better performance as the modules would help each other. Surprisingly, the joint training of $f_\text{T}$ and $f_\text{MSS}$ jointly from scratch results in a lower instrument-wise F1 score,~15.8. Only when we use a pretrained $f_\text{T}$ for 500 epochs and then continue training both $f_\text{T}$ and $f_\text{MSS}$ jointly, do we get a higher F1 score, 24.7.
We hypothesize two reasons for this phenomenon. First, we believe that it is due to the noisy output $\hat{Y}^i_\text{T}$ generated from $f_\text{T}$ in the early stage of joint training which confuses the $f_\text{MSS}$. It in turn causes a wrong gradient being backpropagated to the $f_\text{T}$ model. The same pattern can be observed from the Source-to-Distortion Ratio (SDR) of the $f_\text{MSS}$ which will be discussed in the next sub-section. Second, multi-task training often results in a lower performance than training a model for a single task, partly due to the difficulty in balancing multiple objectives. Our model might be an example of such a case.

When comparing the conditioning strategy, the difference between the summation and the concatenation modes is very subtle. The summation mode outperforms the concatenation mode by only 0.1 F1 score in terms of piece-wise F1 score; while the concatenation mode outperforms the summation mode by 0.001 in terms of flat F1 as well as the instrument-wise F1. We believe that the model has learned to utilize piano rolls to enhance the mel spectrograms. And therefore, summing both the piano rolls after the linear projection with the Mel spectrograms is enough to achieve this objective, and therefore no obvious improvement is observed when using the concatenation mode.

Compared to the existing methods, our model (24.8) outperforms Omnizart~\cite{8682605} (1.9) by a large margin in terms of Instrument-wise F1. This result proves that adding a control mechanism to alter the model behavior when predicting different musical instruments without changing the model architecture is more scalable than expanding the output channels of the models as the number of musical instruments increases. 
%
%

Jointist (58.4), however, did not outperform the MT3 model (76.0) on the note-wise F1 score. This difference may come from two aspects. First, MT3 is based on the Transformer architecture, which outperforms earlier architectures such as ConvRNN, on which our model is based. Second, MT3 leverages multiple transcription datasets. These two aspects are independent from the main modification of Jointist (instrument-conditioning and joint training with source separation). Therefore, it would be possible in the future to train a system that has the merits of both of these frameworks. To test the robustness of proposed framework, we also evaluate the performance by using $f_{IR}$ to generate $I^i_\text{cond}$ (prefix `I-' in Table~\ref{tab:transcription}). We obtain a similar flat F1 scores as our previous experiments, which implies that $f_{IR}$ is good enough to pick up all of the necessary instruments for the transcription. Because false positive piano rolls and false negative piano rolls have undefined F1 scores, we are unable to report the piece-wise and instrument-wise F1 scores. The end-to-end transcription samples generated by Jointist are available in the supplementary material\footnotemark[2].
%

\subsection{Source Separation}
\label{sec:discussion_ss}
Table~\ref{tab:source_separation} shows the SDR for different source separation models. ``S only'' is standalone training of $f_\text{MSS}$. the models ``T sum S'' and ``T cat S'' have a jointly trained $f_\text{T}$ and $f_\text{MSS}$ with $\hat{Y}^i_\text{T}$ and $X_\text{STFT}$ being summed and concatenated respectively. It can be seen that with the help of the transcription module $f_\text{T}$, Joinist is able to achieve a higher SDR. Similar to the discussion in Section~\ref{sec:discussion_transcription}, there is only a minor difference in instrument/piece/source SDR between the ``sum'' mode and ``cat'' (2.01/3.72/3.50~dB vs. 1.92/3.75/3.52~dB). The SDR for each instrument is available in the supplementary material\footnotemark[2].
Our experimental results in both Section~\ref{sec:discussion_transcription} and Section~\ref{sec:discussion_ss} show that the joint training of $f_T$ and $f_\text{MSS}$ helps the other modules to escape local minimum and achieve a better performance compared to training them independently. The source separation samples produced by Jointist are available in the supplementary material\footnotemark[2].

We also explore the upper bound of source separation performance when the ground truth $Y^i_\text{T}$ is used (last row of Table~\ref{tab:source_separation}). When $f_\text{MSS}$ has access to an accurate transcription result, its SDR can be greatly improved. This shows that AMT is an important MIR task that could potentially benefit other downstream tasks such as music source separation.

\section{Applications of Jointist}
\label{sec:applications}


\subsection{Downbeat, Chord, and Key Estimations}

%
It is intuitive to believe that symbolic information is helpful for beat and downbeat tracking, chord estimation, and key estimation. 
Literature has indicated that the timing of notes is highly related to beats \cite{tempobeatdownbeat:book}. Downbeats, on the other hand, correspond to bar boundaries and are often accompanied by harmonic changes \cite{durand2016feature}. 
The pitch information included in piano rolls offers explicit information about the musical key and chords \cite{pauws2004musical,humphrey2015four}. It also provides strong cues for modeling downbeats since they are correlated with harmonic changes. Given these insights, we attempt to 
apply our proposed hybrid representation to improve the performance of these tasks.

For audio processing, we use trained, 6-channel harmonic spectrograms (128 frequency bins) from \cite{won2020data}. We simplify our piano rolls into 2 channel (instrument index 0-37 as channel 0 and index 38 as channel 1). We use a 1-D convolution layer to project the piano rolls into the same frequency dimension as the spectrograms. The hybrid representation is a concatenation of the spectrograms and piano rolls, i.e., 8-channnel. 

SpecTNT~\cite{lu2021spectnt}, a strong Transformer architecture, was chosen for modeling temporal musical events in audio recordings~\cite{hung2022modeling,wang2022catch}. 
The timestamp and label annotations are converted into temporal activation curves for the learning targets for SpecTNT as done in \cite{lu2021spectnt,hung2022modeling}. 
Table~\ref{tab:specTNT_config} in Appendix summarizes our SpecTNT configurations for the three tasks. 

We train beat and downbeat tracking tasks jointly following \cite{hung2022modeling}, but focus on downbeat evaluation, which is more challenging than beat tracking~\cite{bock2020deconstruct,durand2016feature}. We consider 24 classes of major and minor triad chords for chord estimation, and 24 classes of major and minor keys for key estimation, plus a ``none'' class for both tasks. The key and chord tasks are trained and evaluated separately. 

For downbeat tracking, we use 7 datasets: Ballroom~\cite{krebs2013rhythmic}, Hainsworth~\cite{hainsworth2004particle}, SMC~\cite{gouyon2006computational}, Simac~\cite{holzapfel2012selective}, GTZAN~\cite{marchand2015swing}, Isophonics~\cite{mauch2009omras2}, and RWC-POP~\cite{goto2002rwc}. We use either Isophonics or RWC-POP for evaluation while the remaining 6 datasets are used for training. For chord estimation, we use Billboard~\cite{burgoyne2011expert}\footnote{Due to missing audio files for Billboard, we collected them manually from the Internet.}, Isophonics, and RWC-POP. We use either Isophonics or RWC-POP for evaluation while the remaining 2 datasets are used for training. For key estimation, we use Isophonics for evaluation and Billboard for training. 


Table~\ref{tab:result} presents the evaluation results for each task. We report the frame-wise Major/Minor score for chord, F-measure score for downbeat tracking, and the song-level accuracy.
We observe that the model with the hybrid representation can consistently outperform that with spectrogram only across three tasks. This can be attributed to the advantage of piano rolls that provide explicit rhythmic and harmonic information to SpecTNT, which is frequency-aware (i.e., not shift invariant along the frequency axis). 

\begin{table}
\small
\centering
\begin{tabular}{cc|ccc}
\begin{tabular}{@{}c@{}}\textbf{Eval} \\ \textbf{Data}\end{tabular} & \textbf{Input} & \begin{tabular}{@{}c@{}}\textbf{Downbeat} \\ (F1)\end{tabular}    & \begin{tabular}{@{}c@{}}\textbf{Chord}\\(MajMin)\end{tabular}          & \begin{tabular}{@{}c@{}}\textbf{Key} \\ (Acc)\end{tabular}     \\
\toprule
\multirow{2}{*}{A} 
& Audio Only & 0.728 & 0.798  & 0.728 \\
& Hybrid & \textbf{0.746} & \textbf{0.812}   & \textbf{0.752}  \\
 \midrule
\multirow{2}{*}{B} 
  & Audio Only & 0.620  &  0.766  & - \\
  & Hybrid & \textbf{0.663} & \textbf{0.785}   & -  \\
 \bottomrule
\end{tabular}
\caption{Comparison of with and without pianoroll representation on each task (DB: downbeat). A and B represent the dataset Isophonics and RWC-POP, respectively}
\label{tab:result}
\end{table}

\subsection{Music Classification}
We experimented if the piano roll is useful in music classification. MagnaTagATune dataset~\cite{law2009evaluation} is a widely used benchmark in automatic music tagging research. We used $\approx$21k tracks with top 50 tags following previous works introduced in \cite{won2020evaluation}.

Since music classification using symbolic data is a less explored area, we design a new architecture that imitates the music tagging transformer~\cite{won2021semi}. The size of piano roll input is (B, 39, 2913, 88), where B is the batch size, 39 is the number of instruments, 2913 is the number of time steps, and 88 is the number of MIDI note bins. The CNN front end has 3 convolutional blocks with residual connections, and the back end transformer is identical to the music tagging transformer~\cite{won2021semi}. The area under the receiver operating characteristic curve (ROC-AUC) and the area under the precision-recall curve (PR-AUC) are reported as evaluation metrics.

As shown in Table~\ref{tab:mtat}, the piano roll only model does not outperform the existing audio-based approaches. This is somewhat expected since we lose timbre-related information which is crucial in music tagging. However, after a careful analysis of tag-wise metrics, we conjecture strong dependency of the model on instrument information, which indicates that our hybrid approach in this section is under-optimized. This is based on the following observation.

\input{tables/mtat}

\vspace{-0.2cm}
\begin{itemize}[leftmargin=*]
    \item The performance is comparable when an instrument tag is one of our 39 instruments (e.g., cello, violin, sitar), or a genre is highly correlated with our 39 instruments (e.g., rock, techno).
    \vspace{-0.2cm}
    \item The performance drops when a tag is not related to the 39 instruments (e.g., female vocal, male vocal), or the tag is related to acoustic characteristics that cannot be captured by piano rolls (e.g., quiet)
\end{itemize}
\vspace{-0.2cm}

We further experimented with whether a hybrid model with a mid-level fusion (concatenating the audio embeddings and the piano roll embeddings before the transformer back end) improves the performance. As shown in Table~\ref{tab:mtat}, the hybrid model reported performance gain by taking advantage of acoustic features, but could not outperform audio-only models. However, this is a preliminary result and a more thorough experiment and optimization would be required to draw a conclusion.

\section{Conclusion}
\label{sec:conclusion}
In this paper, we introduced Jointist, a framework for transcription, instrument recognition, and source separation. 
Jointist was designed so that different but related tasks can help each other and improve the overall performance. In the experiments, we showed that jointly trained music transcription and music source separation models are beneficial to each other, leading to a highly practical music transcription model. We also showed that transcription results can be used together with spectrograms to improve the model performance of downbeat tracking, chord, and key estimation. While such a hybrid representation did not improve music tagging, the piano roll alone was enough to produce a decent music tagging result.

In the future, Jointist can be improved in many ways, e.g., replacing the ConvRNN with Transformers as done in \cite{hawthorne2021sequence}. We also hope more attempts to be made to use symbolic representations, complementary to audio representations, for progress towards a more complete music analysis system.



\bibliography{ISMIRtemplate}

\clearpage
\begin{appendices}

\input{tables/MIDI_map}
\input{tables/ablation_full}

\begin{table*}
\centering
\begin{tabular}{c|ccc}
Task                   & Input length       & ($k$, $d$)           & ($h_k$, $h_d$)     \\
\midrule
Downbeat      & 6 seconds                  & (128, 96)                 & (4, 8)   \\
Chord   & 12 seconds     & (64, 256)         &   (8, 8)           \\
Key   & 36 seconds                 & (128, 32)      & (8, 4)                 \\
 \bottomrule
\end{tabular}
\caption{SpecTNT parameters we used in each task, where $k$ and $d$ denote spectral and temporal feature dimensions; while $h_k$ and $h_d$ represent the number of heads for the spectral and temporal Transformer encoders, respectively.}
\label{tab:specTNT_config}
\end{table*}

\end{appendices}
\end{document}

%% file: tables/instrument_recognition.tex
\begin{table}[tp]
\center
\small
\begin{tabular}{c|cc|cc}
      & \multicolumn{2}{c|}{\textbf{mAP}}                              & \multicolumn{2}{c}{\textbf{F1}}                               \\
\toprule
\begin{tabular}{@{}c@{}}\#Layers \\ (\#Parameters)\end{tabular} & Macro     & Weighted       & Macro      & Weighted       \\\midrule
1 (78.0M)     & 71.8                  & 91.6          & 61.5                & 85.4          \\
2 (78.8M)    & 72.1                  & 92.1          & 65.0                 & 86.2          \\
3 (79.6M)    & 73.7               & 92.2          & 63.7                 & 86.9          \\
4 (80.3M)      & \textbf{77.4}  & \textbf{92.6} & \textbf{70.3} & \textbf{87.6} \\
\bottomrule
\end{tabular}
\caption{The accuracy of instrument recognition by the number of transformer layers. }
\label{tab:inst}
\end{table}


%% file: tables/transcription.tex
\begin{table}[tp]
\centering
\small
\begin{tabular}{c|cc|cc|cc}
{
}         & \multicolumn{2}{c|}{\textbf{Flat F1}} & \multicolumn{2}{c|}{\textbf{Piece. F1}} & \multicolumn{2}{c}{\textbf{Inst. F1}} \\
\toprule
Model                     & N.     & N\&O    & N            & N\&O      & N             & N\&O     \\
\midrule
\cite{8682605}  & 26.6    & 13.4           & 11.5           & 6.30             & 4.30            & 1.90            \\
\cite{gardner2021mt3}  & 76.0    & 57.0           & N.A.           & N.A.             & N.A.            & N.A.            \\
T              & 57.9    & 25.1           & 59.7           & 27.2             & 48.7            & 23.2            \\
iT              & 58.0    & 25.4           & 59.7           & 27.2             & 48.7            & 23.2            \\\hline
pTS(s)          & \textbf{58.4}    & 26.2           & 61.2           & \textbf{28.5}    & 50.6            & 24.7            \\
ipTS(s)          & \textbf{58.4}    & \textbf{26.5}           & N.A           & N.A    & N.A            & N.A            \\
TS(s)            & 47.9    & 18.6           & 45.7           & 18.4             & 34.9            & 15.8            \\\hline
pTS(c)         & \textbf{58.4}    & 26.3           & \textbf{61.3}  & 28.6             & \textbf{50.8}   & \textbf{24.8}   \\
TS(c)             & 47.7    & 18.6           & 46.0           & 18.0             & 35.5            & 16.2            \\
\bottomrule
\end{tabular}
\caption{Transcription accuracy by training methods. `T' and `S' specifies the trained modules, e.g., `T` indicates that only the transcription module is trained (no source separation). The prefix `p-' represents that the transcription module is pretrained. The prefix `i-' represents that $f_{IR}$ is used to obtain the instrument conditions. (s) and (c) indicate whether the piano rolls are summed or concatenated, respectively, to the spectrograms.}
\label{tab:transcription}
\end{table}

%% file: tables/source_separation.tex
\begin{table}[tp]
\centering
\small
\begin{tabular}{c|ccc}
\textbf{Model}                   & \textbf{Instrument}       & \textbf{Piece}           & \textbf{Source}     \\
\toprule
S only     & 1.52                  & 3.24                 & 3.03   \\
T \textit{sum} S   & \textbf{2.01}     & 3.72        & 3.50              \\
T \textit{cat} S   & 1.92                 & \textbf{3.75}      & \textbf{3.52}                 \\
Upper bound T + S  & {4.06}         & {4.96}        & {4.81}       \\
\bottomrule
\end{tabular}
\caption{Source-to-Distortion Ratio (SDR) for different models. T represents transcriptor, S represents separator. \textit{sum} and \textit{cat} are different ways to merge the piano rolls with the spectrograms. The 4th row shows the upper bound when we use the ground truth piano rolls for source separation.}
\label{tab:source_separation}
\end{table}


%% file: tables/ablation.tex
\begin{table}[tp]
\small
\centering
\begin{tabular}{clccc}
\textbf{Feature merge}                    & \textbf{Model}           & \textbf{Inst.}                  & \textbf{Piece}                & \textbf{Source}              \\\toprule
\multirow{4}{*}{sum}
                                  & T+S   & 1.86                  & 3.55                 & 3.32                 \\\cline{2-5}
                                  & pTS & 2.01                 & 3.72                 & 3.50                           \\\cline{2-5}
                                  & TS (STE) & 1.45                  & 3.31                 & 3.10                       \\\cline{2-5}
                                  & pTS (STE) & 1.99                  & 3.42                 & 3.24                           \\\hline
                             
\multirow{4}{*}{concat}
                                  & TS    & 1.80                  & 3.53                 & 3.31                           \\\cline{2-5}
                                  & pTS   & 1.92                  & \textbf{3.75}        & \textbf{3.52}                  \\\cline{2-5}
                                  & TS (STE) & 2.01                  & 3.58                 & 3.37                        \\\cline{2-5}
                                  & pTS (STE)& \textbf{2.13}         & 3.66                 & 3.46                         \\
                                  \bottomrule
\end{tabular}
\caption{Results for the ablation study with different feature merging and inferring methods. STE stands for straight-Through estimator; pT indicates that a pretrained weight is used to initialize the transcriber $f_T$ for training.}
\label{tab:ablation_source}
\end{table}

%% file: tables/mtat.tex
\begin{table}[tp]
\centering
\small
\begin{tabular}{cl|cc}

\textbf{Input} & \textbf{Model}        & \textbf{ROC-AUC}       & \textbf{PR-AUC}\\
\toprule
\multirow{4}{*}{\begin{tabular}{@{}c@{}}Audio \\ Only\end{tabular} } & \cite{pons2019musicnn}     & 0.9106           & 0.4493\\
& \cite{lee2017sample}    & 0.9058           & 0.4422\\
& \cite{won2020evaluation}    & 0.9129                  & 0.4614     \\
& \cite{won2020data}    & 0.9127           & 0.4611\\
\midrule
{\begin{tabular}{@{}c@{}}Piano Roll Only\end{tabular} } & Transformer   & 0.8938           & 0.4063\\
Hybrid & Transformer     & 0.9090           & 0.4400\\

\bottomrule
\end{tabular}
\caption{Music tagging results on MagnaTagATune Dataset. The audio-only baseline systems are MusiCNN \cite{pons2019musicnn}, Sample-level CNN \cite{lee2017sample}, Short-chunk ResNet \cite{won2020evaluation}, and Harmonic CNN \cite{won2020data}. 
}
\label{tab:mtat}
\end{table}

%% file: tables/MIDI_map.tex
\begin{table*}
\small
\centering
\begin{tabular}{clllc}
\toprule
MIDI Index & MIDI Instrument & MIDI Program     & Our Mapping          & Our Index \\
\midrule
0-3     & Piano                    & Grand/Bright/Honky-tonk Piano    & Piano                & 0         \\\hline
4-5     & Piano                    & Electric Piano 1-2        & Electric Piano       & 1         \\\hline
6     & Piano                    & Harpsichord             & Harpsichord          & 2         \\\hline
7     & Piano                    & Clavinet                & Clavinet             & 3         \\\hline
8-15     & Chr. Percussion       & \makecell[l]{Celesta, Glockenspiel, Music box,\\Vibraphone, Marimba, Xylophone,\\Tubular Bells, Dulcimer}                 & Chr. Percussion & 4         \\\hline
16-20    & Organ                 & \makecell[l]{Drawbar, Percussive,\\Rock, Church,\\Reed Organ}           & Organ                & 5         \\\hline
21    & Organ                    & Accordion               & Accordion            & 6         \\\hline
22    & Organ                    & Harmonica               & Harmonica            & 7         \\\hline
23    & Organ                    & Tango Accordion         & Accordion            & 6         \\\hline
24-25    & Guitar                & \makecell[l]{Acoustic Guitar (nylon, steel)} & Acoustic Guitar      & 8         \\\hline
26-31    & Guitar                & \makecell[l]{Electric Guitar (jazz, clean, muted,\\overdriven, distorted, harmonics)}  & Electric Guitar      & 9         \\\hline
32-39    & Bass                  & Acoustic/Electric/Slap/Synth Bass          & Bass                 & 10        \\\hline
40    & Strings                  & Violin                  & Violin               & 11        \\\hline
41    & Strings                  & Viola                   & Viola                & 12        \\\hline
42    & Strings                  & Cello                   & Cello                & 13        \\\hline
43    & Strings                  & Contrabass              & Contrabass           & 14        \\\hline
44    & Strings                  & Tremolo Strings         & Strings              & 15        \\\hline
45    & Strings                  & Pizzicato Strings       & Strings              & 15        \\\hline
46    & Strings                  & Orchestral Harp         & Harp                 & 16        \\\hline
47    & Strings                  & Timpani                 & Timpani              & 17        \\\hline
48-51    & Ensemble              & Acoustic/Synth String Ensemble 1-2       & Strings              & 15        \\\hline
52-54    & Ensemble              & Aahs/Oohs/Synth Voice              & Voice                & 18        \\\hline
55    & Ensemble                 & Orchestra Hit           & Strings              & 15        \\\hline
56    & Brass                    & Trumpet                 & Trumpet              & 19        \\\hline
57    & Brass                    & Trombone                & Trombone             & 20        \\\hline
58    & Brass                    & Tuba                    & Tuba                 & 21        \\\hline
59    & Brass                    & Muted Trumpet           & Trumpet              & 19        \\\hline
60    & Brass                    & French Horn             & Horn                 & 22        \\\hline
61-63    & Brass                 & Acoustic/Synth Brass           & Brass                & 23        \\\hline
64-67    & Reed                  & Soprano, Alto, Tenor, Baritone Sax             & Saxophone            & 24        \\\hline
68    & Reed                     & Oboe                    & Oboe                 & 25        \\\hline
69    & Reed                     & English Horn            & Horn                 & 22        \\\hline
70    & Reed                     & Bassoon                 & Bassoon              & 26        \\\hline
71    & Reed                     & Clarinet                & Clarinet             & 27        \\\hline
72    & Pipe                     & Piccolo                 & Piccolo              & 28        \\\hline
73    & Pipe                     & Flute                   & Flute                & 29        \\\hline
74    & Pipe                     & Recorder                & Recorder             & 30        \\\hline
75-79    & Pipe                  & \makecell[l]{Pan Flute, Blown bottle, Shakuhachi,\\Whistle, Ocarina}                & Pipe                 & 31        \\\hline
80-87    & Synth Lead            & Lead 1-8         & Synth Lead           & 32        \\\hline
88-95    & Synth Pad             & Pad 1-8                 & Synth Pad            & 33        \\\hline
96-103    & Synth Effects        & FX 1-8                  & Synth Effects        & 34        \\\hline
104-111   & Ethnic               &  \makecell[l]{Sitar, Banjo,\\ Shamisen, Koto,\\ Kalimba, Bagpipe,\\ Fiddle, Shana}                    & Ethnic               & 35        \\\hline
112-119   & Percussive           & \makecell[l]{Tinkle Bell, Agogo, Steel Drums,\\Woodblock, Taiko Drum, Melodic Tom,\\Synth Drum}             & Percussive           & 36        \\\hline
120-127   & Sound Effects        & \makecell[l]{Guitar Fret Noise, Breath Noise,\\Seashore, Bird Tweet, Telephone Ring,\\Helicopter, Applause, Gunshot}       & Sound Effects        & 37        \\\hline
128   & Drums                   & Drums                  & Drums               & 38        \\\hline
\bottomrule
\label{tab:MIDI_map}
\end{tabular}
\caption{The instrument mapping used in our experiments. Our mapping is less detailed than the MIDI Program Number, but it is finer than the MIDI Instrument code, thus resulting in 39 different instruments.} 
\end{table*}

%% file: tables/ablation_full.tex
\begin{table*}[ht]
\centering
\begin{tabular}{clcccccc}   & & \multicolumn{3}{c}{{ \textbf{Full length SDR}}} & \multicolumn{3}{c}{{ \textbf{10s SDR}}} \\\toprule
feature merge                   & Model           & inst                  & piece                & source               & inst               & piece              & source            \\\hline\hline
\multirow{8}{*}{sum}        & T+S (binary)       & 1.35                  & 3.22                 & 2.97                 & 3.14               & 5.32               & 4.31              \\\cline{2-8}
                                  & T+S (posterior)    & 1.86                  & 3.55                 & 3.32                 & 3.47               & 5.60               & 4.54              \\\cline{2-8}
                                  & preT+S  (binary)    & 0.20                  & 2.75                 & 2.44                 & 1.86               & 5.34               & 4.11              \\\cline{2-8}
                                  & preT+S  (posterior) & 2.01                 & 3.72                 & 3.50                 & \textbf{3.69}      & 5.86               & 4.81              \\\cline{2-8}
                                  & T+S+STE  (binary)   & 1.30                  & 3.30                 & 3.06                 & 2.40               & 5.13               & 4.16              \\\cline{2-8}
                                  & T+S+STE  (posterior) & 1.45                  & 3.31                 & 3.10                 & 2.41               & 4.96               & 4.07              \\\cline{2-8}
                                  & preT+S+STE  (binary)    & 1.67                  & 3.30                 & 3.06                 & 3.29               & 5.36               & 4.41              \\\cline{2-8}
                                  & preT+S+STE  (posterior) & 1.99                  & 3.42                 & 3.24                 & 3.26               & 5.17               & 4.29              \\\hline\hline
                             
\multirow{8}{*}{concat}     & T+S  (binary)       & 1.28                  & 3.20                 & 2.95                 & 2.72               & 5.06               & 4.20              \\\cline{2-8}
                                  & T+S  (posterior)    & 1.80                  & 3.53                 & 3.31                 & 3.21               & 5.50               & 4.51              \\\cline{2-8}
                                  & preT+S  (binary)    & -0.04                 & 2.63                 & 2.31                 & 1.86               & 5.44               & 4.09              \\\cline{2-8}
                                  & preT+S  (posterior) & 1.92                  & \textbf{3.75}        & \textbf{3.52}        & 3.50               & \textbf{6.03}      & \textbf{4.90}     \\\cline{2-8}
                                  & T+S+STE  (binary)   & 1.89                  & 3.60                 & 3.37                 & 3.63               & 5.70               & 4.66              \\\cline{2-8}
                                  & T+S+STE  (posterior) & 2.01                  & 3.58                 & 3.37                 & 3.59               & 5.53               & 4.55              \\\cline{2-8}
                                  & preT+S+STE  (binary)    & 1.72                  & 3.47                 & 3.22                 & 3.18               & 5.78               & 4.61              \\\cline{2-8}
                                  & preT+S+STE  (posterior) & \textbf{2.13}         & 3.66                 & 3.46                 & 1.86               & 5.44               & 4.09              \\\hline\hline
\multirow{8}{*}{spec patch} & T+S  (binary)       & 1.74                  & 3.42                 & 3.20                 & 3.22               & 5.19               & 4.37              \\\cline{2-8}
                                  & T+S  (posterior)    & 1.79                  & 3.46                 & 3.24                 & 3.35               & 5.30               & 4.43              \\\cline{2-8}
                                  & preT+S  (binary)    & 0.38                  & 2.72                 & 2.41                 & 1.91               & 3.60               & 3.39              \\\cline{2-8}
                                  & preT+S  (posterior) & 1.91                  & 3.61                 & 3.40                 & 3.58               & 5.52               & 4.65              \\\cline{2-8}
                                  & T+S+STE  (binary)   & 1.74                  & 3.46                 & 3.24                 & 3.50               & 5.28               & 4.42              \\\cline{2-8}
                                  & T+S+STE  (posterior) & 1.94                  & 3.48                 & 3.27                 & 3.48               & 5.15               & 4.30              \\\cline{2-8}
                                  & preT+S+STE  (binary)    & 1.71                  & 3.47                 & 3.23                 & 3.28               & 5.40               & 4.53              \\\cline{2-8}
                                  & preT+S+STE  (posterior) & 2.01                  & 3.58                 & 3.38                 & 3.57               & 5.30               & 4.48              \\\hline\hline
\end{tabular}
\caption{Full version of Table~\ref{tab:ablation_source}. `spec patch' represents the model variation where the final learned mask is applied to the summation of the piano roll features and the spectrograms instead of the original spectrograms. We also experimented with two forms of piano rolls: posteriorgram and binary. Posteriorgram provides the probability of the existence of a note in the input audio to the model, which outperforms the model that uses binary version of the piano roll.}
\label{tab:ablation_source_full}
\end{table*}